 \documentclass[smallabstract,smallcaptions]{dccpaper}

\usepackage{epsfig}
\usepackage{citesort}
\usepackage{amsmath}
\usepackage{amssymb}
\usepackage{color}
\usepackage{url}
\usepackage{pdfpages}

\newlength{\figurewidth}
\newlength{\smallfigurewidth}

\begin{document}

\title
{\large
\textbf{ED: Perceptually tuned Enhanced Compression Model}
}

\author{%
Pierrick Philippe, Théo Ladune, Stéphane Davenet and Thomas Leguay\\[0.5em]
{\small\begin{minipage}{\linewidth}\begin{center}
\begin{tabular}{ccc}
Orange Innovation \\
Cesson Sevigne, France\\
\url{firstname.lastname@orange.com}
\end{tabular}
\end{center}\end{minipage}}
}

\Support{Submitted to DCC 2024 in the course of the Challenge on Learned Image Compression (CLIC)\\
© 2024 IEEE.  Personal use of this material is permitted.  Permission from IEEE must be obtained for all other uses, in any current or future media, including reprinting/republishing this material for advertising or promotional purposes, creating new collective works, for resale or redistribution to servers or lists, or reuse of any copyrighted component of this work in other works.}

\maketitle
\thispagestyle{empty}

\begin{abstract}

This paper summarises the design of the candidate ED for the Challenge on
Learned Image Compression 2024. This candidate aims at providing an anchor based
on conventional coding technologies to the learning-based approaches mostly
targeted in the challenge.

The proposed candidate is based on the Enhanced Compression Model (ECM)
developed at JVET, the Joint Video Experts Team of ITU-T VCEG and ISO/IEC MPEG.

Here, ECM is adapted to the challenge objective: to maximise the perceived
quality, the encoding is performed according to a perceptual metric, also the
sequence selection is performed in a perceptual manner to fit the target bit per
pixel objectives.

The primary objective of this candidate is to assess the recent developments in
video coding standardisation and in parallel to evaluate the progress made by
learning-based techniques. To this end, this paper explains how to generate
coded images fulfilling the challenge requirements, in a reproducible way,
targeting the maximum performance.

\end{abstract}

\Section{Introduction}

In the Joint Video Experts Teams ITU-T VCEG (Q6/16) and ISO/IEC MPEG (JTC 1/SC
29/WG 5) are studying future video coding technology with a compression
capability that significantly exceeds that of the current Versatile Video Coding
(VVC) standard~\cite{9503377}. The developments are made on a common software
platform called ECM (Enhanced Compression Model).
\newline

In September 2023, the 10th version of ECM, ECM10 has been released. In Intra
mode coding, it is reported~\cite{ECM10} that ECM10 achieves 12.5\% of bit rate reduction
compared to the VTM~\cite{VTM} (the reference software implementation of VVC)
under identical coding conditions. These results are reported for video content,
i.e. for 4:2:0 sampling formats. Consequently, ECM appears to be a good
candidate technology for image compression as well.
\newline

The image compression task of for the 6th Challenge on Learned Image Compression
consists in compressing a corpus of images at three bitrates: 0.075, 0.150 and
0.300 bit per pixels. The winners are chosen based on a human rating task
implying the need to subjectively optimise the image coding process.
\newline

This paper proposes to adapt ECM to the challenge conditions. As such, this paper:

\begin{itemize}
\item provides the appropriate conversion for image coding;
\item select the encoding parameters to maximise the coding quality;
\item provides selection means in order to reach the three target bit rates.
\end{itemize}

The paper is structured as follows. The first part presents a concise
description of the ECM coding scheme, describes the colour conversion scheme and
explain the appropriate coding parameters for ECM. The coded image selection
process is detailed and the results in objective performance are reported. A
final section derives some conclusion and further perspectives.

\Section{Encoding process}

This section gives some description on the ECM technology and explain the coding
configuration adopted for this challenge.

The ECM is essentially based on the VVC technology. VVC tools are refined and
extended to further improve coding efficiency. A detailed description of the ECM
version 10 tools is available in~\cite{ECM}.

\SubSection{ECM Core coding and RGB to YUV conversion}

ECM improves all the essential Intra coding tools. For instance, intra
prediction is enhanced through decoder side mode derivation or directional
planar mode. Also template matching tools further extend the prediction scheme.
The transformation stages are also improved thanks to additional learned primary
and secondary transforms: their sizes are extended and non-separable transforms
are generalised.
\newline

As the ECM is not the final design of a standard, its operating range is
currently limited to the core video coding application: Although the VTM has
support for 4:4:4 image and video sources, most of the tools provided in ECM to
further improve the compression efficiency are currently only adapted to 4:2:0
formats. Consequently, ECM is only applicable to sub-sampled chroma formats.
\newline

For this submission the RGB image content, sampled in 4:4:4 is therefore
converted to YUV 4:2:0 10 bits, in two standard steps:

\begin{itemize}
\item YUV conversion according to the BT.709 colour space~\cite{InternationalTelecommunicationUnion2015i};
\item Downsampling of the U and V components, the Lanczos filter is selected here;
\end{itemize}

As some sequences have odd width or height, padding of one pixel is necessary
prior to chroma downsampling to ensure that the chroma resolution is exactly
half of the luma resolution.
\newline

Since the objective of this challenge is to optimise the subjective quality a
perceptually driven optimization method is selected for the encoding process. A
quantization parameter adaptation (QPA), was proposed for VVC~\cite{K0206}. The
quantisation adaptation strives to optimise a weighted PSNR metric, called
XPSNR.
\newline

XPSNR is a weighted PSNR motivated by the fact that coding artifacts are often
only perceivable in specific regions. Distortion in highly textured areas are
less visible than in low-contrast regions. Consequently, the distortion mean squared error
is weighted in a block-based manner to take into account the local
contrast of the image. Due to the block-based approach, XPSNR can be directly
turned into a rate distortion cost in a block-based coding scheme. Initialy
proposed for VVC (VTM), it is adapted here for the ECM.
\newline

Given the correlation of the XPSNR metric with numerous subjective
assessments~\cite{9054089}, this metric is selected in this paper in order to
optimise the perceptual quality.
\newline

Table~\ref{tab:command} gives the command line used for ECM encoding.

\begin{table*}

\caption{ECM Software coding configuration.\label{tab:command}}
\footnotesize
\centering
\begin{tabular}{lp{0.5\textwidth}}
\hline
  Option &
  Description \\
\hline
\texttt{--InputFile } & Selects the input file\\
\texttt{--BitstreamFile } & Indicates the bistream file\\
\texttt{--SourceWidth } & Selects the video width\\
\texttt{--SourceHeight } & Selects the video height\\
\texttt{--InputBitDepth=10 } & The source is processed on 10 bits\\
\texttt{-c encoder\_intra\_ecm.cfg} & Selects the basic coding configuration, for Intra format\\
\texttt{--Level=6.2 } & Selects a profile and level compatible with higher resolution images\\
\texttt{--QP qp} & Specifies the base value of the quantization parameter \\
\texttt{--SliceChromaQPOffsetPeriodicity=1} & Periodicity for inter slices that use the slice-level chroma QP offsets \\
\texttt{--PerceptQPA=1} & Applies perceptually optimised QP adaptation \\
\texttt{--ConformanceWindowMode=1} & Automatic padding of the input source to fit the ECM coding unit sizes  \\
\hline
\end{tabular}

\end{table*}

\Section{Bitstream selection}

For the challenge an average number 0.075, 0.150 and 0.300 bit per pixel targets are defined for the image coding
challenge. These targets are not defined individually for each image but for the
entire set of pictures. For the validation set, the average picture resolution is
around 3 million pixels, and the total allowed byte budget is respectively
823660, 1647321 and 3294643 bytes for the three targets.
\newline

How the bit budget is distributed among the sequences is let to the proponent.
\newline

For this candidate the individual bit budget per image is selected based on the
VMAF~\cite{vmaf} metric. The VMAF is widely used and one the most reliable
metric today. Here, for the three target, the selection is made in order to
maximise the minimum VMAF among the images.
\newline


The following process is performed:

\begin{enumerate}
\item source RGB to YUV 4:2:0 conversion on 10 bits, with padding when necessary;
\item ECM encoding within a QP range (here the range is qp 27 to 47) with QPA (see the command line table~\ref{tab:command});
\item ECM decoding;
\item decoded YUV to RGB 4:4:4 conversion, and one pixel cropping for odd resolutions;
\item VMAF measurement;
\item iteratively allocate more bits to the sequence with the worse VMAF;
\item the iteration stops when the bit budget is reached.
\end{enumerate}

In other words, the sequence with the lower VMAF is allocated more bits in a
progressive manner until the bit budget constraint is reached.

\Section{Results and comparison with VTM}

As the ECM operates internally in 4:2:0 sampling format, one can wonder whether
this sampling has an impact especially for the higher bit rate range (0.300
bpp). In order to confirm that the ECM maintains its advantage with respect to
VVC, the validation pictures are also encoded using VTM (version 19) with the
procedure described before, omitting the chroma downsampling operation.
\newline

\begin{figure}[t]
\begin{center}
   \includegraphics[width=0.8\linewidth]{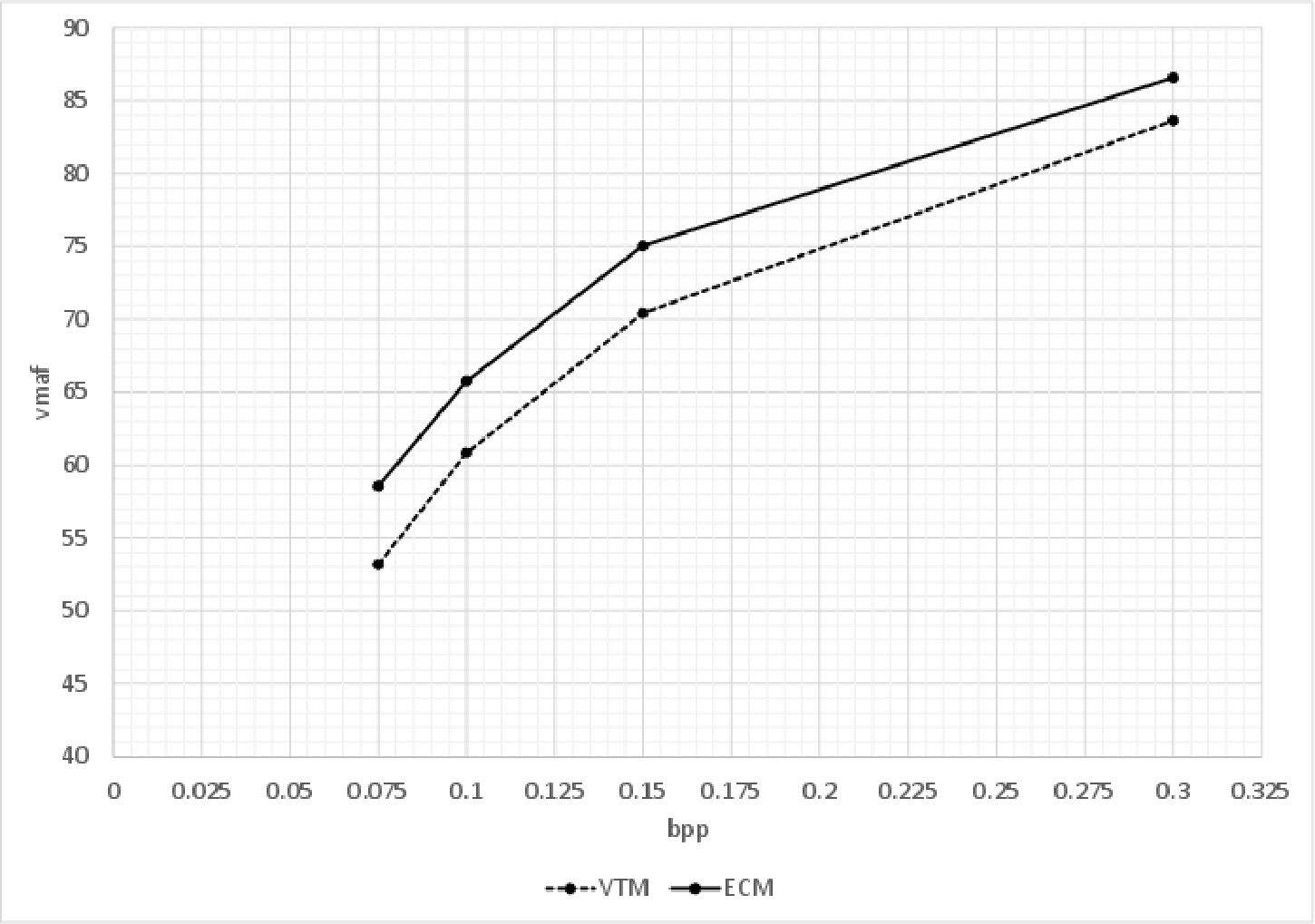}
\end{center}
   \caption{Comparison of the VMAF scores obtained by the ECM submission and the VTM. Despite the chroma subsampling, ECM in 4:2:0 outperforms the VTM operating in 4:4:4}
\label{fig:coding}
\end{figure}

Figure~\ref{fig:coding} shows the relative performance of the ECM candidate in
4:2:0 compared to the VTM operating in 4:4:4: the ECM candidate consistently
outperforms the VTM according to the VMAF evaluation despite its operation in the
subsampled domain. Roughly 5 grades on the VMAF scale largely differentiate the two
coding schemes. This can be turned into a relative bit rate gain of about 19\%
using the Bjøntegaard delta measure in favor of the ECM.
\newline

This confirms that the ECM solution seems to be an appropriate candidate for this challenge.
\newline

Table~\ref{tab:metrics} summarises some objective measures for this submission.
The PSNR and MS-SSIM are the one reported on the validation leaderboard. The
VMAF metrics are also reported, through the average over the sequences and the
lowest VMAF among the 30 images (which is the criterion maximised here).
\newline

\begin{table*}
\centering
\begin{tabular}{l c c c c c c}
\hline
	 bpp &   {Data size} 	&   average VMAF 	&  worse VMAF 	&    PSNR 	&    MS-SSIM	\\
	0.075 &   821963 		&   60.563       	&  58.556 		&    28.006 &    0.936 	\\
	0.150 &   1637153 		&   76.311 			&  75.045 		&    30.436 &    0.963 	\\
	0.300 &   3261508 		&   87.406 			&  86.572 		&    33.076 &    0.978 	\\
\hline
\end{tabular}

\caption{Objective metrics for the 3 challenge bit rates for the ED candidate}
\label{tab:metrics}
\end{table*}

\Section{Conclusion}

This paper provides a perceptually tuned version of ECM, the technology beyond
the VVC standard developed at JVET. To provide a set of images targeting
subjective evaluation, the encoding is performed using a QP adjustment method.
Also, the bit rate allocated for each image is selected according to the VMAF
metric highly correlated with a perceived quality.

\Section{References}
\bibliographystyle{IEEEbib}
\bibliography{refs}

\end{document}